**An Experiment in Using Virtual Worlds**
**for Scientific Visualization of Self-Gravitating Systems**
Will Meierjurgen Farr, Piet Hut, Jeff Ames, Adam Johnson

**Abstract**

In virtual worlds, objects fall straight down.  By replacing a few lines of code to include Newton's gravity, virtual world software can become an N-body simulation code with visualization included where objects move under their mutual gravitational attraction as stars in a cluster. We report on our recent experience of adding a gravitational N-body simulator to the OpenSim virtual world physics engine.  OpenSim is an open-source, virtual world server that provides a 3D immersive experience to users who connect using the popular "Second Life" client software from Linden Labs.  With the addition of the N-body simulation engine, which we are calling NEO, short for N-Body Experiments in OpenSim, multiple users can collaboratively create point-mass gravitating objects in the virtual world and then observe the subsequent gravitational evolution of their "stellar" system.  We view this work as an experiment examining the suitability of virtual worlds for scientific visualization, and we report on future work to enhance and expand the prototype we have built.  We also discuss some standardization and technology issues raised by our unusual use of virtual worlds.

All too often visualization is an afterthought in physics simulation. Producing proper visualization tools is complicated (often more complicated than producing the simulation to be visualized) and uninteresting from the standpoint of physics. Here we present a simple example of a visualization system for N-body gravitational dynamics built on top of the virtual world system OpenSim (The OpenSim Developers, 2008) an open-source version of the software used in Second Life.

From the point of view of an astrophysicist dealing with gravitational N-body simulations, virtual worlds such as OpenSim are N-body simulators, with two extra features: a surprisingly elaborate graphics module and a bug in the equations of motion. As to the latter: whereas objects should attract each other via Newton's inverse-square law of gravity, objects in OpenSim fall straight down. However, that "bug" is easily fixed. We have done so and we discuss our first results in this paper.

Our visualization system, NEO, or N-Body Experiments in OpenSim, runs within the OpenSim server. We allow users connected to the server to designate objects within the virtual world as "physical." Physical objects interact gravitationally as point masses. A small amount of modified code in the OpenSim physics engine tracks the motion of physical objects under their collective gravitational forces. OpenSim displays their motion along with the other objects and users in the virtual world. OpenSim provides facilities for users to communicate with each other using text or voice, allows them to trade files or in-world objects, and allows easy creation and manipulations of in-world objects. With a few hundred lines of code added the OpenSim physics engine, we have a 3D collaborative visualization system for experiments with point-mass gravitating systems.

**The OpenSim Platform**

OpenSim is an open-source C# program which implements the Second Life virtual world server protocol. Running the popular Second Life client software from Linden Labs, users can connect to a computer or grid of computers running the OpenSim server and enter a virtual world. Within this world users can share media—text, pictures, and video—interact with each other via text chat and voice communication, and create and share 3D objects in the world itself. The interaction occurs in real-time: at each moment, every logged-in user views the current 3D state of the server world. Different users can manipulate this state simultaneously.

The 3D, interactive nature of the OpenSim virtual world makes it an ideal substrate for collaborative visualization of scientific results and simulations. The "hard" parts of collaborative visualization unrelated to the science—interactivity, 3D display, controls, etc.—are handled by the pre-existing OpenSim engine, leaving scientists free to focus on the best way to represent their scientific data within the virtual world.

Our gravitational simulation code, which we will discuss more fully in the next section, lives in the physics engine of OpenSim. In "vanilla" OpenSim, the physics engine is responsible for tracking the positions and velocities of primitive objects and users, implementing effects such as falls, tumbles, and collisions. The server requests the positions and velocities of all objects under its control from the physics engine ten times every second; clients wishing for a larger frame-rate use the velocities of the objects to extrapolate their positions at intermediate points.

## The Newtonian Physics Engine

The physics engine of OpenSim handles the updating of the positions and velocities of all objects and avatars in the virtual world. Though velocity information is not strictly necessary for a client to render a scene, it is used by the client to extrapolate the positions of prims and characters between updates from the server. The standard physics engine is extremely simplistic. Prims are divided into two classes: *physical* and *unphysical*. Physical prims feel the effects of a uniform gravitational field (that is, they fall straight down just as physical objects do on Earth), while unphysical prims simply move in straight lines with constant velocity. Both types of prims can collide with other solid objects.

We have modified the standard physics engine of OpenSim using a plugin. Server administrators can select to replace the standard physics engine with our plugin at server-initialization time, region by region. (In OpenSim different servers correspond to different regions in the virtual world; administrators can choose to employ our plugin on a region-by-region basis.) The modified physics engine treats each physical prim as a gravitating point-mass in space; other objects are handled by dispatching to the standard physics engine. We have implemented a variety of integration algorithms for time-advancing the resulting gravitational system in the Newtonian Physics engine: the Hermite algorithm (Makino, 1991) (the default), kick-drift-kick and drift-kick-drift leapfrog, and the GL3 algorithm (Farr & Bertschinger, 2007).[1]

The current simulation is rudimentary. We choose units so that $G = 1$, $M = \sum_i m_i = 1$, and the total energy of the system is $E = -\frac{1}{4}$. (These are the so-called "standard units" (Heggie & Mathieu, 1986). In these units, the average inverse pair-wise separation in an equal-mass system of bodies is $\left\langle \frac{1}{r_{ij}} \right\rangle = 1$. The pair-wise gravitational potential is softened to prevent extreme two-body interactions that destroy the accuracy of the integrator:

$$V(r) = \frac{m_1 m_2}{\sqrt{r^2 + \varepsilon^2}},$$

where $\varepsilon = \frac{4}{N}$; $N$ is the number of bodies in the system. The softening ensures that the maximum two-body interaction potential, $V_{max} \approx \frac{m_1 m_2}{\varepsilon} = \frac{1}{4N}$, is of the same order as the typical equipartition kinetic energy of a body $\langle T \rangle \approx \frac{|E|}{N} = \frac{1}{4N}$ in equilibrium. Softening prevents any individual encounter between two bodies from changing the trajectories of either body too much, greatly simplifying the implementation of the simulation.

---

[1] For an introduction to writing N-body code, see Hut & Makino, 2009, especially "Moving Stars Around." For general background concerning self-gravitating systems, see Heggie & Hut, 2003 and for background concerning N-body algorithms, see Aarseth, 2003.

In these natural units, the typical time for a body to cross from one side of a system to the other is of order unity.  The size of the system is also of order unity.  Using these dimensionless units, there is no need for conversion between "server time" and N-body time and "server length" and N-body length, so the user can see the system evolve on realistic time- and length-scales.  (For example, the two-body relaxation timescale for a system with $N \sim 30$ is about $t_{cr} \frac{N}{0.1 \ln N} \sim 250$ seconds of real time.)  We can simulate about $N \sim 50$ bodies in this fashion on typical modern desktop hardware before the server cannot keep up with the necessary frame rates to the connected clients.  Though 50 bodies is small by modern simulation standards, such a system is sufficient to illustrate most of the physical behaviors important in larger systems before core collapse—evaporation, two-body relaxation, mass segregation, etc.  We could increase the maximum number of bodies that can be simulated by not demanding that the simulation and display remain synchronized at the cost of introducing significant complexity in the code.

## Examples

This section, as an example, presents screenshots of an interaction in OpenSim simulating about 30 bodies starting from a cold initial condition.  In Figure 1, the avatar sets up an initial condition by creating a group of objects (by holding "shift" while dragging the movement bars, a pre-existing group of bodies can be copied).  Figure 2 captures the system a moment after the avatar has selected the "Physical" box; very soon after (on the free-fall timescale, which is of order one second in the simulator), in Figure 3, the two groups of bodies in the initial condition quickly collapse, forming the two clumps visible in the figure.  The simulation ends after about a minute of simulator time (a few tens of crossing times) in Figure 4, with a collapsed, nearly-spherical cluster and a few almost-ejected stars in loose orbits.

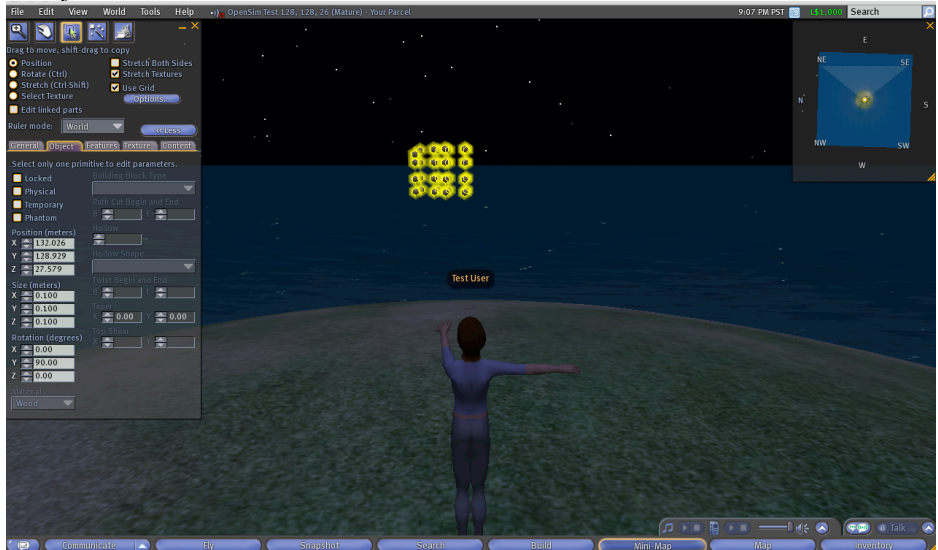

**Figure 1: Establishing an initial condition.**

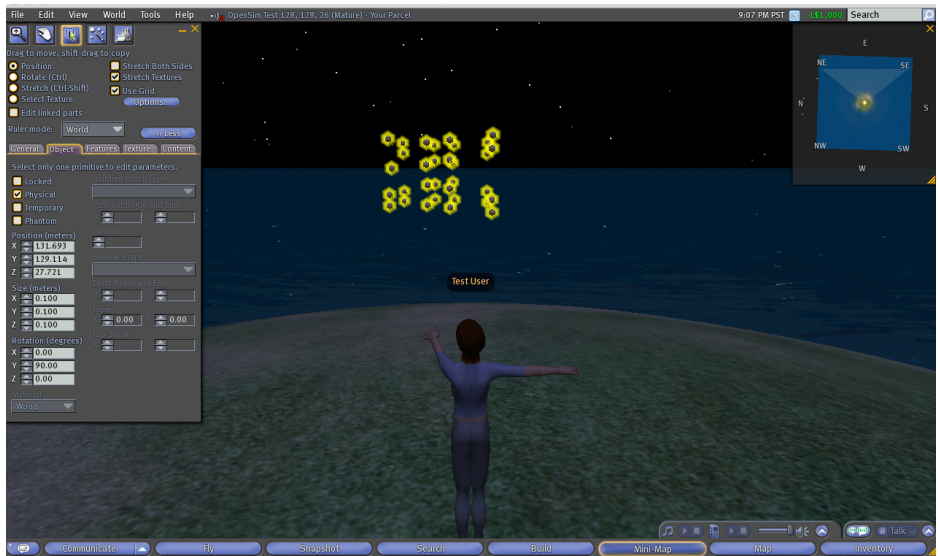

**Figure 2: When the user selects the "Physical" box, the system is scaled into standard units and the simulation begins.**

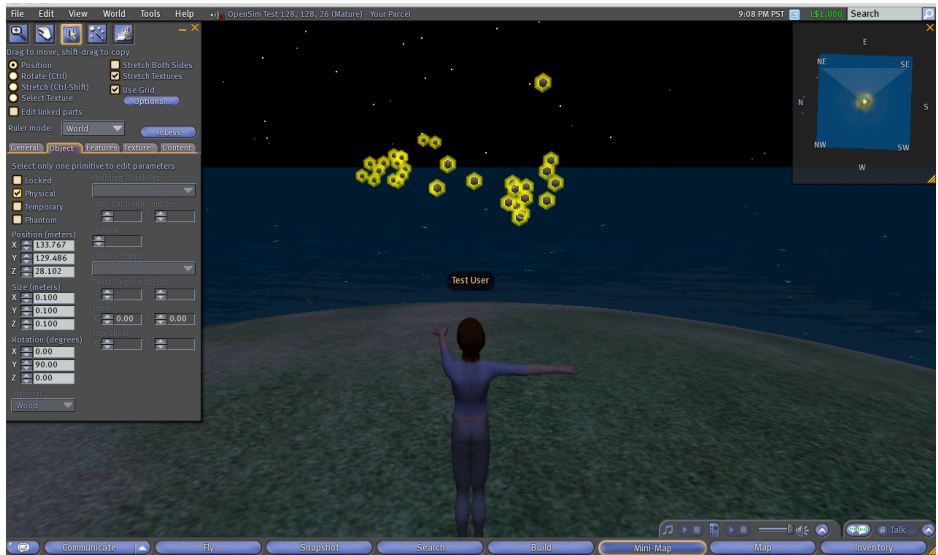

**Figure 3: Two groups of bodies in the initial condition collapse into two clumps on the free-fall timescale.**

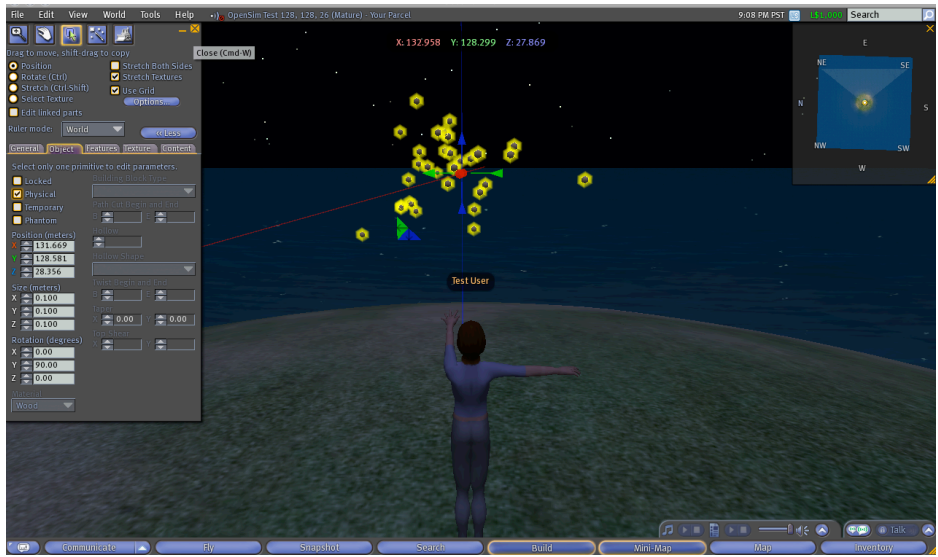

**Figure 4: After a few tens of crossing times (a few tens of seconds real time), the system settles down into a spherical cluster.  Some nearly-ejected stars can be seen orbiting the central mass of the cluster.**

Though this example shows only one avatar in view on a remote "desert island," a similar simulation could, in principle, take place anywhere on an OpenSim grid, and any user present could collaborate to construct the initial conditions, discuss the outcome with other avatars, save data from the simulation, etc.

### Limitations and Future Work

This section discusses some of the limitations of the current simulation engine, and highlights future work that promises to resolve them.

The biggest limitation of the current engine is the size of the simulations it can run.  A system of 50 bodies is sufficient to *illustrate* the phenomena that are important in physically relevant simulations, but to *study* physical systems, simulations must be much larger.  The cost of a simulation of a quasi-equilibrium cluster of gravitating bodies over an evolutionary timescale grows approximately as $N^3$; it is not reasonable to expect to perform physically relevant simulations in real-time on a virtual world server.  To address this limitation, a collaboration between the National Institute for Informatics and the National Astronomical Observatory of Japan (The AstroSim Project, 2009) is preparing a visualizer that allows users to re-play a simulation conducted on a more powerful computer inside a virtual world; this harnesses the speed advantages of specialized hardware (e.g. the GRAPE-DR Project, 2008) for the simulation, and the collaborative advantages of virtual worlds for the visualization.

Even with the size limitations inherent in the server-based simulation engine we describe, it can still be an useful tool for education and enhanced understanding of the microphysics of self-gravitating systems.  It would be more useful for these purposes, however, if it had the capability to start systems in more varied initial conditions.

**Sys Admin 2/10/70 3:42 AM**
**Comment:** Write this out. I'm assuming they mean "National Institute for Informatics" and National Astronomical Observatory of Japan.

Currently, the system only permits cold (i.e. zero velocity) initial conditions for systems of bodies that must be constructed by hand in the virtual world. Ideally, it would permit the specification of arbitrary initial conditions (perhaps via a notecard) and the quick creation of various analytically determined distributions of stars. Work is in progress to permit this.

Finally, further control over the simulation would be desireable. At a minimum, one should be able to pause and restart the simulation easily—currently, this requires de-selecting the physical property for all bodies in the simulation). It would also be nice to add discrete physical events by hand (i.e. insert or remove stars from the simulation, change the mass of stars which "go supernova," etc). Work is in progress to add a simplified control panel that appears in the virtual world.

The work reported here has been carried out during the summer of 2008 in Tokyo at the National Astronomical Observatory of Japan. Since then, we have continued our work in collaborations that involve several co-workers at the National Institute for Informatics, also in Tokyo, and other co-workers whom we met with regularly in the Meta Institute for Computational Astrophysics (MICA) in Second Life (see http://www.mica-vw.org/; Djorgovski, et al., 2009a & 2009b; Nakasone, et al., 2009). Most recently, we are conducting a weekly workshop to discuss the use of virtual worlds for stellar dynamics, in collaboration between MICA and Kira (http://www.kira.org/; see http://www.kira.org/index.php?option=com_content&task=view&id=124&Itemid=154).

### Issues for Technology and Standardization

Our unusual use case and implementation techniques for the N-body physics engine raise a number of issues related to technology and standardization. In this section, we attempt to identify some of these issues. We discuss these issues in the context of our N-body physics engine, but they would be relevant for any scientific simulation conducted in a virtual world.

The data in our simulation are unusual for a virtual world. Instead of complicated, unmoving structures, we have simple structures executing complicated motion. How can we store the history of the bodies' motion in the virtual world? Would it be possible to represent that history itself as an object in the virtual world? Could avatars trade N-body systems with each other? What about transfer of systems from one virtual world to another?

If we wish to visualize multiple N-body systems, for example, while teaching a class, we need a way to ensure they don't interfere with each other. We may also wish to verify that a simulation has really been isolated during its run, without the rest of the world providing additional influences on the motions of the stars. This may call for a way to isolate different parts of a virtual world from each other for a time to minimize the effects of one on the other.

The simulations we are running can be arbitrarily demanding on the server CPU (to simulate $N$ bodies takes time proportional to $N^2$). This allows for the possibility of an inadvertent overload on the server, which could in practice resemble a denial-of-service attack against the virtual world server. Should we just limit the number of objects an user is allowed to create and simulate? Degrade the quality of the simulation dynamically according to server load?  Treat server CPU as a resource that users must request and manage specifically?

Our application is heavily customized, targeting only OpenSim.  Currently, no standard interface exists for writing plugins to modify the behaviors of the common components of virtual worlds.  Should such interfaces be standardized?  What would such a standard look like?  What would be the risks in opening up the infrastructure of the virtual world to external modification?

Similarly, the procedure to install our application in an OpenSim instance is unique.  Could this be standardized?  If standardized, would a common installation procedure apply only to OpenSim, or for other virtual worlds as well?  A standard plugin installation procedure could make these sorts of modifications available to more users, but most OpenSim server administrators are probably sophisticated enough to not be deterred by a slightly customized installation procedure.

As discussed above, the poor scaling of the computational cost of an N-body simulation with the number of bodies probably requires that any large simulations be performed on a different computer than the virtual world server. The results of such simulations can be passed to the server, which can then provide them to the attached clients for presentation.  However, even though the computational load on the server is minimized in this architecture, the data load could be considerable, particularly when many clients demand the simulation data.  Should standards be created allowing a server to refer clients to another source for some of the data they are to display?  What about a standard for clients sharing data among themselves to reduce the load on the server and the external data source?  What are the possible security implications?  This bandwidth problem is not unique to scientific simulation applications for virtual worlds, but such applications often deal with exceptionally large datasets and therefore the problem is relatively more important for these applications.

While a great amount of thought has been put into technology and standards for the typical use of virtual worlds, the types of uses we discuss here are just beginning to be explored.  For the reasons we discuss above, we think that the future holds great promise for the use of virtual worlds as visualization and simulation platforms.  As such uses become more common the issues in this section—and others—must be addressed.

## Conclusion

We have reported on our experience adding a gravitational N-body simulator to OpenSim.  The simulator exists as a modification of the standard physics engine of OpenSim, and is notable for its simplicity.  Nevertheless, the resulting simulation environment can "piggyback" on all the collaboration features of the OpenSim virtual world to provide a multi-user, interactive environment.  We anticipate that this sort of rich collaboration is the future of scientific visualization and we argue that virtual worlds provide an ideal substrate on which to base such visualization systems.  The work described in this paper has barely scratched the surface of the capabilities of such a system, yet provides a compelling example of the suitability of this approach for creating visualization tools by a quick retooling of the existing infrastructure of a virtual world.

**Acknowledgements**

Will Farr and Piet Hut express their thanks to the NAOJ for inviting them as visitors during the summer of 2008, when the work described here was carried out.  We also want to thank Jun Makino for being our host during that period.  In addition, we thank Helmut Prendinger and Ken Muria from NII for several stimulating conversations.

> **Sys Admin 2/10/70 4:03 AM**
>
> **Comment:** Us? I take it WF and PH stand for their names. If so, they need to be written out